\documentclass{article}

\usepackage{graphicx} 
\usepackage{newtxtext,newtxmath}
\usepackage[T1]{fontenc}

\usepackage{amsthm}

\usepackage{url}
\usepackage{hyperref}
\usepackage[LY1]{fontenc}
\usepackage{graphicx}
\usepackage[round]{natbib}

\title{A Deep Learning Neural Network Algorithm for Classification of Eclipsing Binary Light Curves}
\author{B. Ula\c{s}}
\date{June 2023}

\begin{document}

\maketitle
\begin{abstract}
We present an image classification algorithm using deep learning convolutional neural network architecture, which classifies the morphologies of eclipsing binary systems based on their light curves. The algorithm trains the machine with light curve images generated from the observational data of eclipsing binary stars in contact, detached and semi-detached morphologies, whose light curves are provided by Kepler, ASAS and CALEB catalogues. The structure of the architecture is explained, the parameters of the network layers and the resulting metrics are discussed. Our results show that the algorithm, which is selected among 132 neural network architectures, estimates the morphological classes of an independent validation dataset, 705 real data, with an accuracy of 92\%.
\end{abstract}

\section{Introduction}

Deep learning techniques strengthen their solid ground in various areas from art to science every passing day. In the present, countless machine and deep learning applications let researchers achieve faster and more precise results in their studies, as well as changing the daily life. Convolutional neural networks, a specialized architecture in deep learning algorithms using neural networks, give an opportunity to use powerful methods in some processes such as image recognition and classification. The prototype of these networks, necognitron, was proposed by \citet{fuk80}. \citet{lec98} introduced the convolutional networks by remarking their performance on variability in 2D shapes. They also noted the advantages of fast learning in their handwriting experiment. The improvement in both hardware and software technology allows taking giant leaps in the usage and development of convolutional neural networks. For instance, a famous architecture AlexNet \citep{kri12} classified 1.2 million images with an accuracy value of about 85\% in ImageNet computer vision challenge. CoAtNet \citep{dai21} also reached 91\% accuracy by improving the model capacity and introducing the hybrid models.

Eclipsing binary stars are basically stellar systems showing light variations in their light curves due to occultations of the companions' light. Their importance arises from being tools for deriving the crucial stellar parameters precisely, and therefore, allowing the researchers to determine the structure of the stars in realistic estimations. \citet{gui93} remarked that the analyses of their light curve enable us to estimate important parameters like mass, radius, luminosity, effective temperature as well as atmospheric properties. The systems appeared in several morphological types, mainly contact, detached and semi-detached, \citep{kop55,bra05} that can be corresponded to various phases of their evolution. Thus, determining the morphological classes of these stars relieve information about related light parameters as well as the stellar evolution in different circumstances, which opens the doors to understanding the universe better as we know the binary and multiple systems are very common in our galaxy.  

Researchers made efforts in detecting, fitting and classifying the light curves of binary systems using machine and deep learning algorithms. \citet{wyr03} proposed an algorithm using artificial neural networks and detected 2580 binary systems in Large Magellanic Cloud based on the OGLE data \citep{uda98}. \citet{pri08} presented an artificial neural network trained with data points of 33235 light curve samples of detached eclipsing binaries for deriving some physical parameters of eclipsing binary stars selected from several databases. The authors remarked that the success rate of the algorithm is more than 90\% for OGLE and CALEB data sets, respectively. \citet{koc20} evaluate different fitting methods and concluded that machine learning techniques are useful tools for estimating the initial parameters of the binaries. The preliminary results of a systematic classification for the light curve morphologies of eclipsing binaries from TESS \citep{ric15} performing machine learning technique were published by \citet{bir20}. \citet{ula20} suggested a deep learning image classification algorithm for the classification of light curve morphologies of ASAS-SN eclipsing binaries with an accuracy value of 92\%. Lately, \citet{cok21a} introduced a two-class (detached and over-contact) classification based on the 491425 synthetic light curve data generated by {\tt ELISa} software \citep{cok21b}. The authors accomplished 98\% accuracy with their combined deep learning architecture. 

Applications of machine learning techniques on astrophysical data are developing and promise novel results in the area. The potential of the subject motivated us to apply the method to the eclipsing binary light curves. In the following section, we introduce the details and structure of the light curve data used in the study. Sec.~\ref{code} deals with the details of our code and the architecture of the convolutional neural network. The results are discussed and the concluding remarks are given in the last section.

\section{Light Curve Data}\label{lcdata}

The algorithm needs light curve images corresponding to certain morphological classes of binary stars to train the machine and perform the classification. Therefore, we collected real light curve data to construct light curve images of eclipsing binary stars. The data for eclipsing binary stars with known morphological types are provided from three main data sources in this study; Kepler Eclipsing Binary Catalog \citep{kir16}, All Sky Automated Survey \citep[ASAS,][]{poj97} and Catalog and Atlas of Eclipsing Binaries \citep[CALEB, former EBOLA,][]{bra04}.

Kepler light curves were accessed through the Kepler Eclipsing Binary Catalog \citep{kir16}. The authors catalogued some basic properties of 2920 binary systems and indicated a parameter for their morphological classes. The parameter ($c$), introduced by \citet{mat12} using locally linear embedding method, is a classification criterion for contact ($0.7<c<0.8$), detached ($c<0.5$) and semi-detached ($0.5<c<0.7$) binary systems. We collect 1913 binary systems (239 contact,  1253 detached and 421 semi-detached) with corresponding $c$ parameters and used their phase and detrended flux values from the catalogue to construct the light curve images.

ASAS \citep{poj97} variable star database was also used to gather light curve data and morphological classes of the eclipsing binary stars. The variability class of the targets in the catalogue was determined by \citet{poj02} using an approach based on multidimensional parametric space as well as an extended method using certain Fourier coefficients. We were able to collect data and morphological classes of 5907 binary systems through database query service\footnote{\url{http://www.astrouw.edu.pl/asas/}}. The phases for the light curves were calculated by adopting the times of minimum and orbital period values from the ACVS (ASAS Catalog of Variable Stars) list given by the author. The magnitudes were also converted to normalized fluxes by deriving the maximum magnitudes for corresponding light curves for the systems.

The CALEB data were achieved via the catalogue's web page\footnote{\url{http://caleb.eastern.edu}}. The author catalogued light curves and observational properties of 305 individual stars with their morphological classes. Since the catalogue contains light curves in several filters for many stars, the actual number of the data exceeds the above-mentioned value. 1632 light curves from the database were included in our study. The light curve images were constructed by using the phase and flux values given by the catalogue.  

\begin{figure}
\centering
\includegraphics[width=\columnwidth]{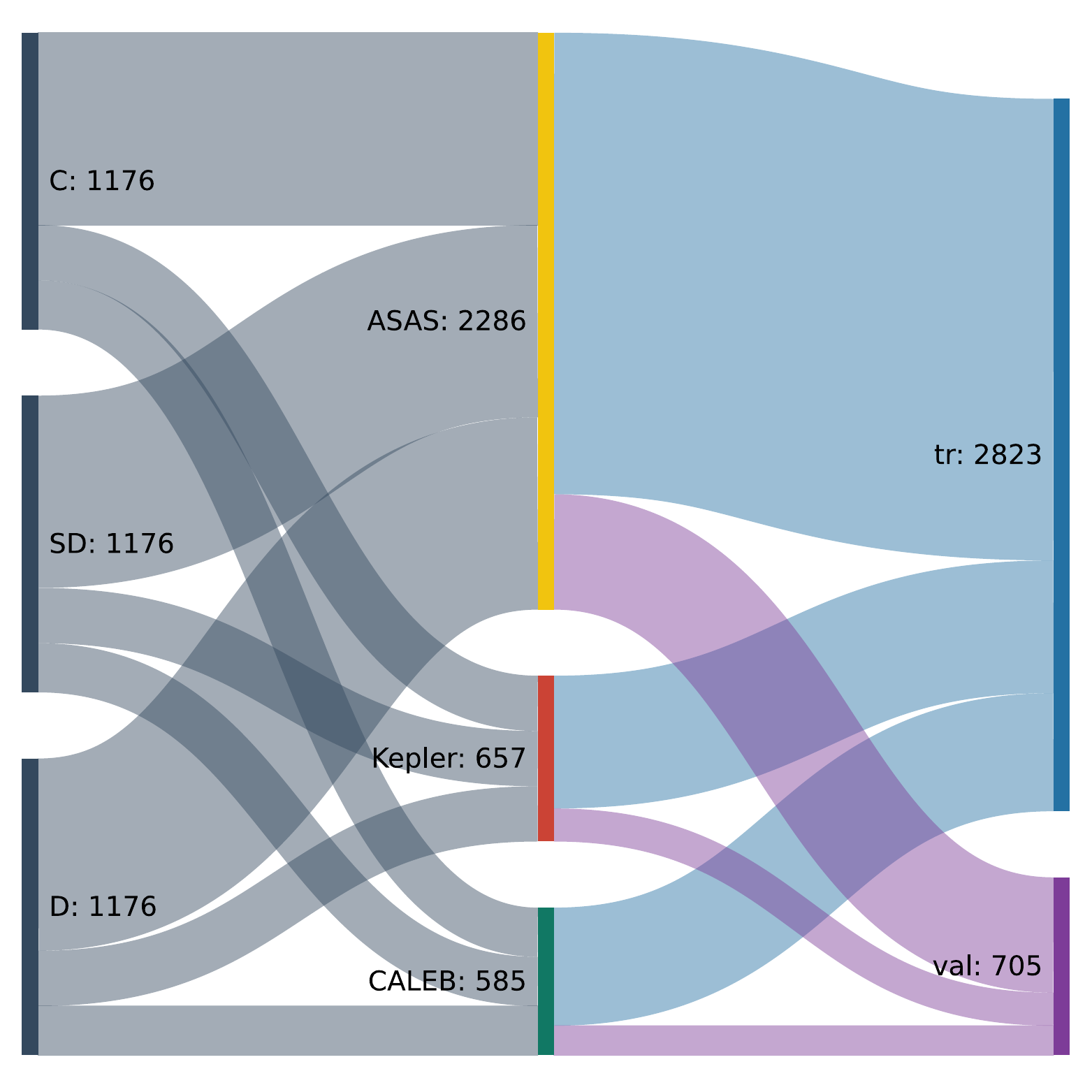}
\caption[]{Sankey diagram showing the distribution of 3528 light curve image data in three nodes; morphology, database and type of the dataset. C, D and SD refer to contact, detached and semi-detached morphologies, respectively. {\bf tr} indicates training set, while {\bf val} remarks the validation data. See text for details. Diagram created using {\tt SankeyMATIC}\protect\footref{san}.}
\label{sankey_plot}
\end{figure}

\begin{figure*}
\centering
\includegraphics[width=\columnwidth]{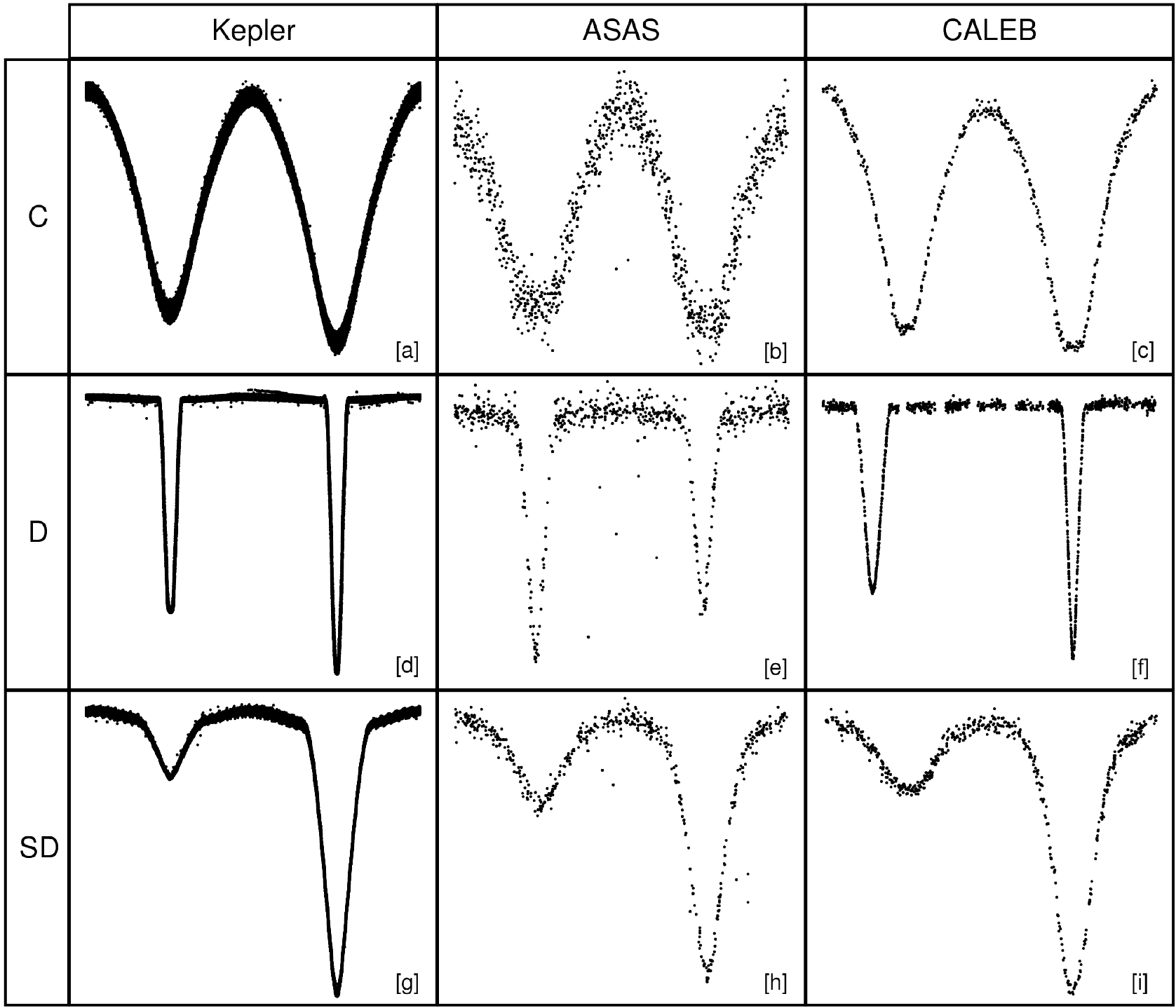}
\caption[]{Selected images from the training set generated by using the light curve data of KIC~12458133 (a), ASAS~065227-5524.6 (b), V572~Cen (c), KIC~06545018 (d), ASAS~075602-4454.8 (e), QX~Car (f), KIC~03954798 (g), ASAS~101553-6012.9 (h) and TZ~Lyr from three different databases. C, D and SD refer to contact, detached and semi-detached morphologies, respectively. Figure created using {\tt gnuplot}\protect\footref{gnu}.}
\label{samples}
\end{figure*}

The $256 \times 256$ pixels light curve images were generated by plotting the data in $0.25-1.25$ phase interval from the mentioned databases. The total number of data was decreased after eliminating the light curves which $(i)$ show very large scattering, $(ii)$ have very few data points and $(iii)$ do not resemble the light curve of an eclipsing binary system. The incorrect orbital period values, especially in ASAS Catalog, were also responsible for the decrement in the number of light curves. Additionally, the entire dataset from three databases was checked by eye to prevent misclassification. The data in each class were also balanced. Namely, we limit the number of light curve images in each morphological class is to be equal, thus, it is one-third of the total number of data in a given database (e.g. 2286 light curves from ASAS contains 762 images from each individual class; contact, detached and semi-detached). We randomly chose 657, 2286 and 585 images from the final datasets of Kepler, ASAS and CALEB, respectively. Therefore, a total of 3528 light curves from all databases were selected to use for image classification. The number of the light curves in the training set is 2823 while the validation set covers 705 data, about 20\% of the training set, following the Pareto principle \citep{moo97,jur99}. Fig.~\ref{sankey_plot} is a Sankey diagram showing the relation among the morphologies, databases and datasets based on the number of light curves. Nine samples of data with different morphologies in the training set from three databases are illustrated in Fig~\ref{samples}.

\addtocounter{footnote}{1}
\footnotetext{\label{san}\url{https://github.com/nowthis/sankeymatic}}

\addtocounter{footnote}{1}
\footnotetext{\label{gnu}\url{https://github.com/gnuplot}}

\section{Architecture of the Neural Network}\label{code}

A Python \citep{van09} code\footref{bu} was written to set a deep learning neural network algorithm and thus train the machine to classify the light curve images generated from the light curve data. The code contains import procedure of the main and axillary sources in the preamble, that are {\it NumPy} package \citep{har20}, {\it os} module \citep{van20}, {\it pandas} library \citep{mck10, pan20}, {\it Random} module \citep{van20}, {\it TensorFlow} platform \citep{aba15} and {\it Keras} API \citep{cho15} which are needed the processes in the algorithm work. It proceeds with seed fixing for NumPy, Python and TensorFlow to avoid randomness and make the results reproducible, and yet randomness that may arise from the calculations on the Graphics Processing Unit (GPU) still remains. It must be noted that when running the code on a GPU, randomness may alter the results slightly from one run to another due to the parallel operations, as remarked by the Keras team. The problem can be solved by conducting the calculations on a Central Processing Unit (CPU), however, neural networks are computationally expensive, and it takes an extremely long time to achieve the results on a CPU.

The image data in three folders (C: contact, D: detached and SD: semi-detached) were indicated in the code and the total number of the data files inside those directories was commanded to display on the output. The sizes of input images were also defined. We applied data augmentation by adding random Gaussian blur to images in the training dataset. Augmentation enriches information about data by applying certain operations to a given dataset, and it helps prevent overfitting \citep{sho19}. As some augmentation methods (e.g. flip and rotation) may cause changes in the shape of the light curve, we avoided employing further augmentation to our data. Training and validation directories were defined to direct the algorithm to the targeted data which was intended to be dealt with. The data generation process resizes the images to $128 \times 128$ pixels to save computing time and converts to greyscale since colour is not a distinctive feature of our data. In data generation, {\tt {class\_mode}} argument was {\tt {categorical}} which defines 2D {\it one-hot} encoded labels that contain one {\it hot} ({\tt 1}) among all other {\it cold} ({\tt 0}) values \citep{har07}.

The backbone of the algorithm is the sequential model, a stack of layers, which includes several hyperparameters forming the convolutional neural network architecture which consists of convolutional, pooling, fully connected and output layers. Convolutional layers use kernels with the size of $(3,3)$, referring to the size of the convolutional window \citep{cho15}. Rectified Linear Unit (ReLU) function, $f(x)=max(0,x)$, was selected as the activation function for the layers, which assigns 0 for the values smaller than or equal to zero \citep{goo16}. The training is done by using the stochastic gradient descent algorithm \citep{rud16} to achieve the converged result and then ReLU provides relatively more effortless optimization and calculation since it represents the nonlinearities with two linear functions. The convolution operation was done by applying a $L2$ regularization penalty \citep{cor09}. The regularization term is:

\begin{equation}
    \lambda \sum_{i=1}^{N} \omega_i^2
	\label{eq:l2reg}
\end{equation}
where $\lambda$ (=0.001 in our case), $\omega$ and $N$ are the regularization parameter, weight and the number of features, respectively. The term adds the squared weights to the loss function and controls the weights to be relatively small values, thus, preventing the model from overfitting and structural complexity. The {\tt padding} hyperparameter was adjusted to {\tt same}, which guarantees that the feature map is the same size as the input \citep{cho15}. The stride value was left default, $(1,1)$ which corresponds that the filter moves one pixel at a time. We also applied max pooling operation \citep{chr19} with a pool size of $(2,2)$ between convolutional layers. It basically downsamples the input data by taking the maximum values within the pool size. The pooling also helps avoid overfitting and lowers the computation time. The above processes lead to the feature extraction and the next stage, the flattening operation, is necessary since the final convolutional layer does not cover the entire dimension of the input image \citep{sha20}. Flattening converts the data into a 1-dimensional array, the shape which is mandatory to make the algorithm be able to perform the classification. A Dropout layer with a rate of 0.5 follows the flattening, which avoids the model from overfitting, as mentioned by \citet{sri14}. The last steps of the convolutional neural network include fully connected layers where the classification takes place. All the input neurons are connected to the neurons in the present layer at this stage \citep{gro17}, therefore, the dimensionality of the upper Dense layer is equal to the filter number of the last convolutional layer. The dimension was set to 3 in the output layer of the network since we have three classes (contact, detached and semi-detached). Probabilistic distribution was determined using softmax activation function as it is appropriate for multiclass classifications using categorical cross-entropy loss function, which is:

\begin{equation}
    L = -\sum_{i=1}^{n} p(x_i) \log_e(q(x_i))
	\label{eq:loss}
\end{equation}
given by \citet{zho21}, where $ p(x_i)$ and $q(x_i)$ denote real and predicted distributions, and $n$ is the number of classes. Layers from the first convolutional to the second last Dense layer, generally called hidden layers, consist of 5 trainable and 6 nontrainable layers.

The Adam algorithm \citep{kin15} was chosen as the optimizer, which uses the stochastic gradient descent method. Adam is appropriate for multiclass problems and can be adjusted with the learning rate hyperparameter. The learning rate is basically referring to the step size \citep{mur12} in the convergence of the learning process. Tuning this parameter plays an important role in obtaining reliable results during calculations. Large values can result in straying from convergence, while small values may cause taking long times of training. We control the learning process by monitoring the validation loss through {\tt EarlyStopping} callback, which is known to boost the performance of algorithms \citep{yao07}. The arguments of early stopping were arranged to stop training when no decrement in validation loss is observed in 20 consecutive epochs, and therefore, training was prevented to be overfit. Another callback, {\tt ModelCheckpoint}, was also included in the code to save the best model having the maximum validation accuracy in a model file. We compiled our model using cross-entropy loss, as mentioned before, based on accuracy evaluation. The final operation, fitting the model, was done by specifying the number of training samples per iteration ({\tt batch\_size=32}), generators, and the callbacks remarked above. Additionally, in our code, we stored the number of filters in convolutional layers and learning rate values in variables ({\tt l1}, {\tt l2}, {\tt l3}, {\tt l4} and {\tt lrate}) to be able to test various architectures quicker, only by changing the set of variables.

\begin{figure}
\centering
\includegraphics[width=\columnwidth]{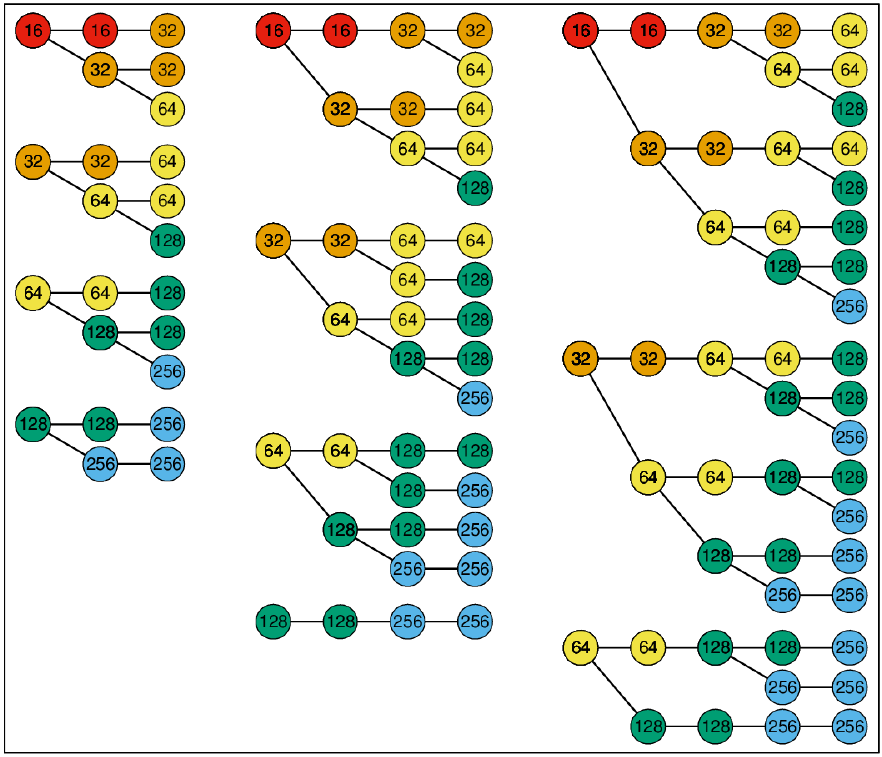}
\caption[]{Schematic representation of 44 different architectures with their filter numbers which are employed with three specific learning rate values ($10^{-3}$, $10^{-4}$, $10^{-5}$) separately, and correspond to 132 different networks.}
\label{allarc_plot}
\end{figure}

\begin{figure*}
\centering
\includegraphics[width=\columnwidth]{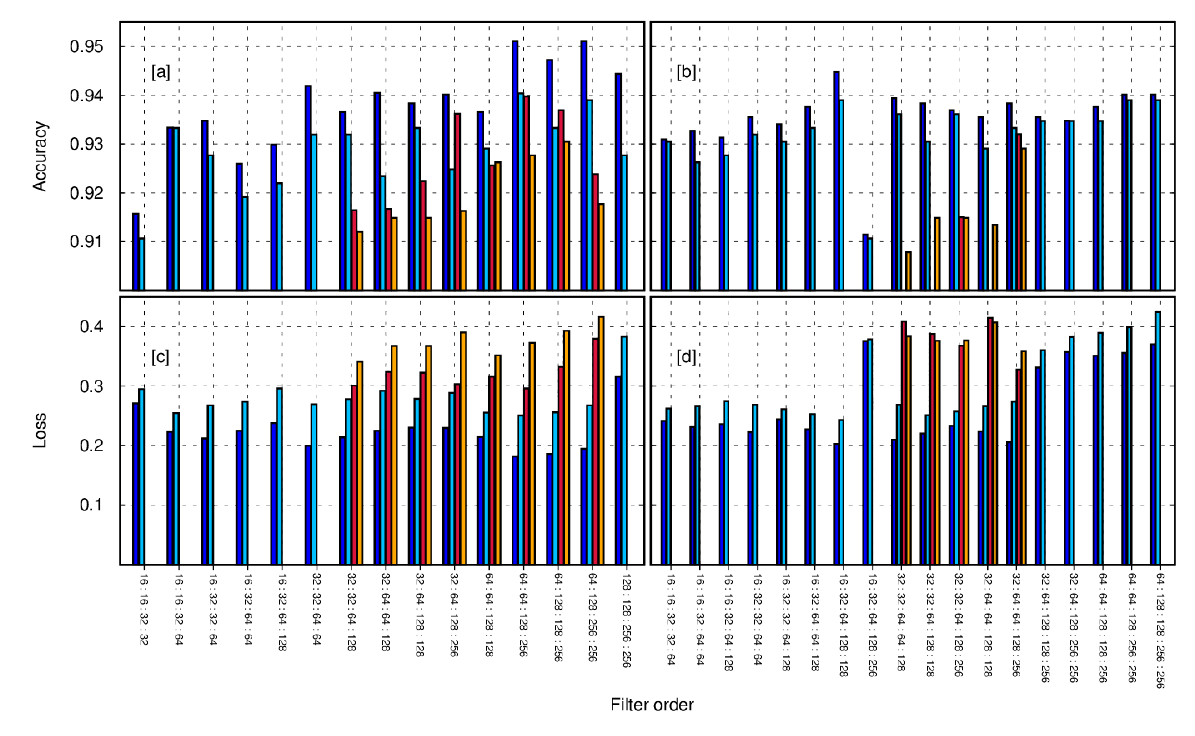}
\caption[]{Final Accuracy (a and b) and Loss (c and d) values for different architectures whose validation accuracies are larger than 0.9. Filter orders refer to the filter numbers of convolutional layers. Blue and light blue refer to the training and validation datasets with the learning rate value of $10^{-4}$. Red and orange bars represent the training and validation results when the learning rate was set to $10^{-5}$.}
\label{acc_loss_plot}
\end{figure*}

The aim of the neural network is to minimize the cross-entropy type loss function and reach the maximum accuracy value. Accuracy is a measure of how model predictions are close to the ones from the real model in all classes, while loss, a cost function, measures whether the predictions give the correct values \citep{sam17}. Specifically, in zero-one loss, 0 and 1 refer to correct and incorrect classifications, respectively. To achieve the best result based on the corresponding values, we employ a total of 132 different convolutional neural network architectures (Fig.~\ref{allarc_plot}) with three different learning rate values ($10^{-3}$, $10^{-4}$, $10^{-5}$) which were run using NVidia T4 GPU accelerator provided by Kaggle\footnote{\url{https://www.kaggle.com}} platform. Although the higher accuracy values were reached in some other architectures, the optimum result was achieved in the network of 5 convolutional layers having 32, 32, 64, 128, 256 filters, respectively (see Sec.~\ref{res}). The final accuracy and loss values for the models with validation accuracies larger than 0.9 are represented in Fig.~\ref{acc_loss_plot}. Models with the learning rate value of $10^{-3}$ are not included in the figure, since their validation accuracy never exceeded 0.9.

\section{Results and Conclusion}\label{res}

\begin{figure}
\centering
\includegraphics[width=\columnwidth]{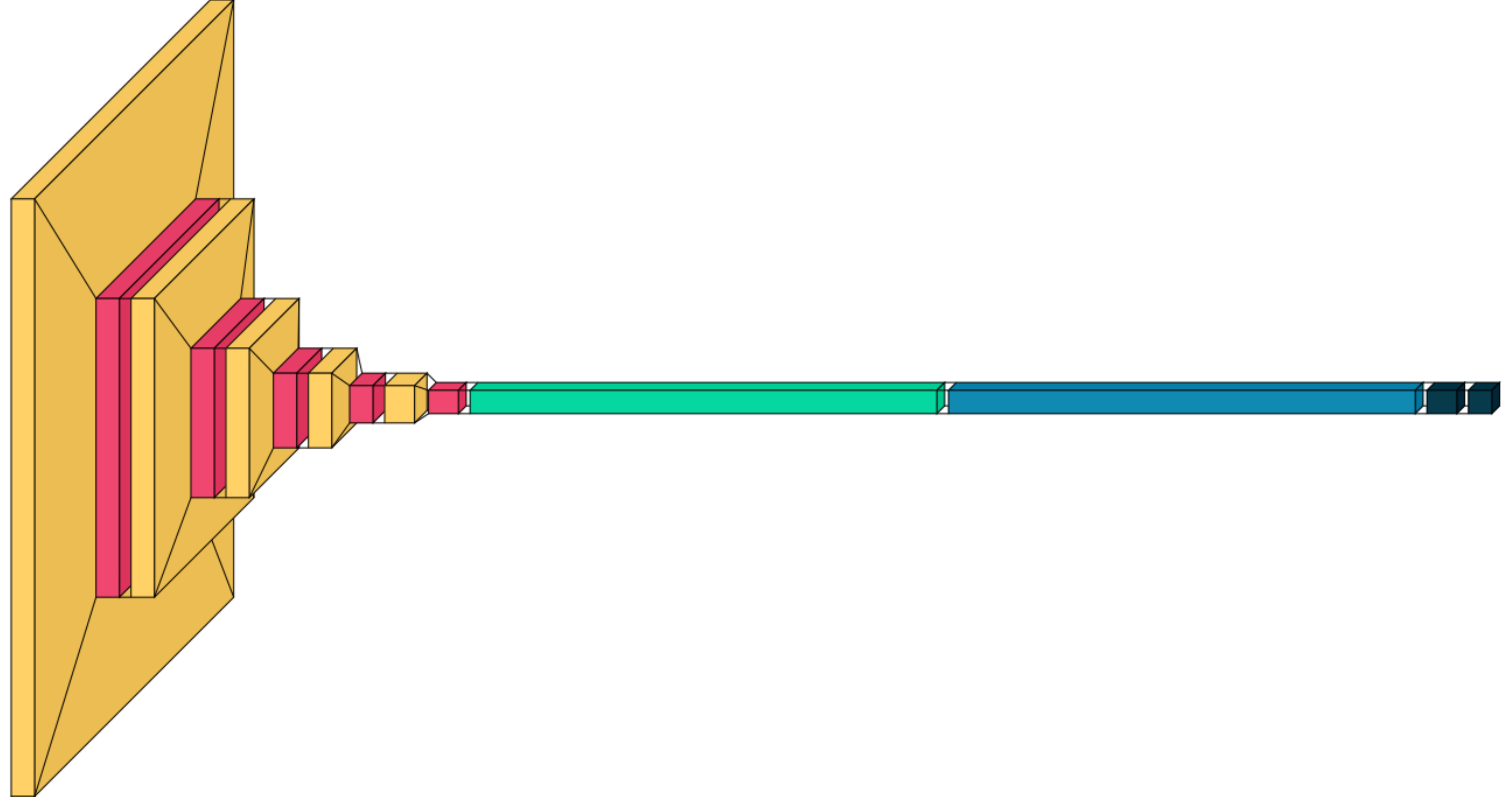}
\caption[]{Visualization of the final neural network architecture. Orange, red, green, blue and black colors refer to Convolutional, Max pooling, Flatten, Dropout and Dense layers. Figure created using {\tt visualkeras for Keras/TensorFlow} \citep{gar20}.}
\label{arc_plot}
\end{figure}

\begin{figure*}
\centering
\includegraphics[width=\columnwidth]{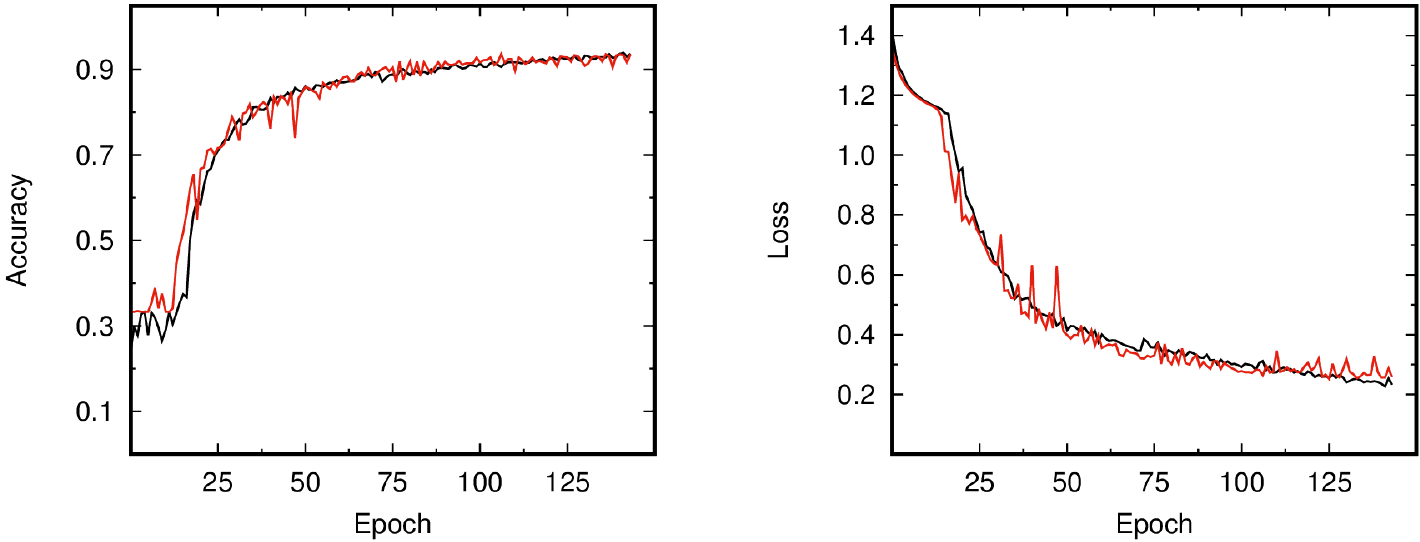}
\caption[]{Learning curves, variation of training (black) and validation (red) accuracy and loss with epoch.}
\label{acc_loss_fin}
\end{figure*}

As mentioned in the previous section, an accuracy of 92\% was achieved in architecture with 5 convolutional layers having 32, 32, 64,128 and 256 filters. The learning rate was $10^{-4}$, and it takes 142 epochs for the architecture to achieve the result, whose training loss (0.233) is slightly lower than validation loss (0.257), and the training and validation accuracies (0.937 and 0.936) are close. Thus, the model can be considered prevented from overfitting and underfitting compared to other models shown in Fig.~\ref{acc_loss_plot}. A visualization of the final architecture is shown in Fig.~\ref{arc_plot}. The learning curves, accuracy and loss values for both training and validation versus epoch, are plotted in Fig.~\ref{acc_loss_fin}. The trends of the curves imply a typical good fit. The KERAS model file containing the final architecture is provided through GitHub repository\footnote{\label{bu}\url{https://github.com/burakulas/ebclass}}.

\begin{figure}
\centering
\includegraphics[width=\columnwidth]{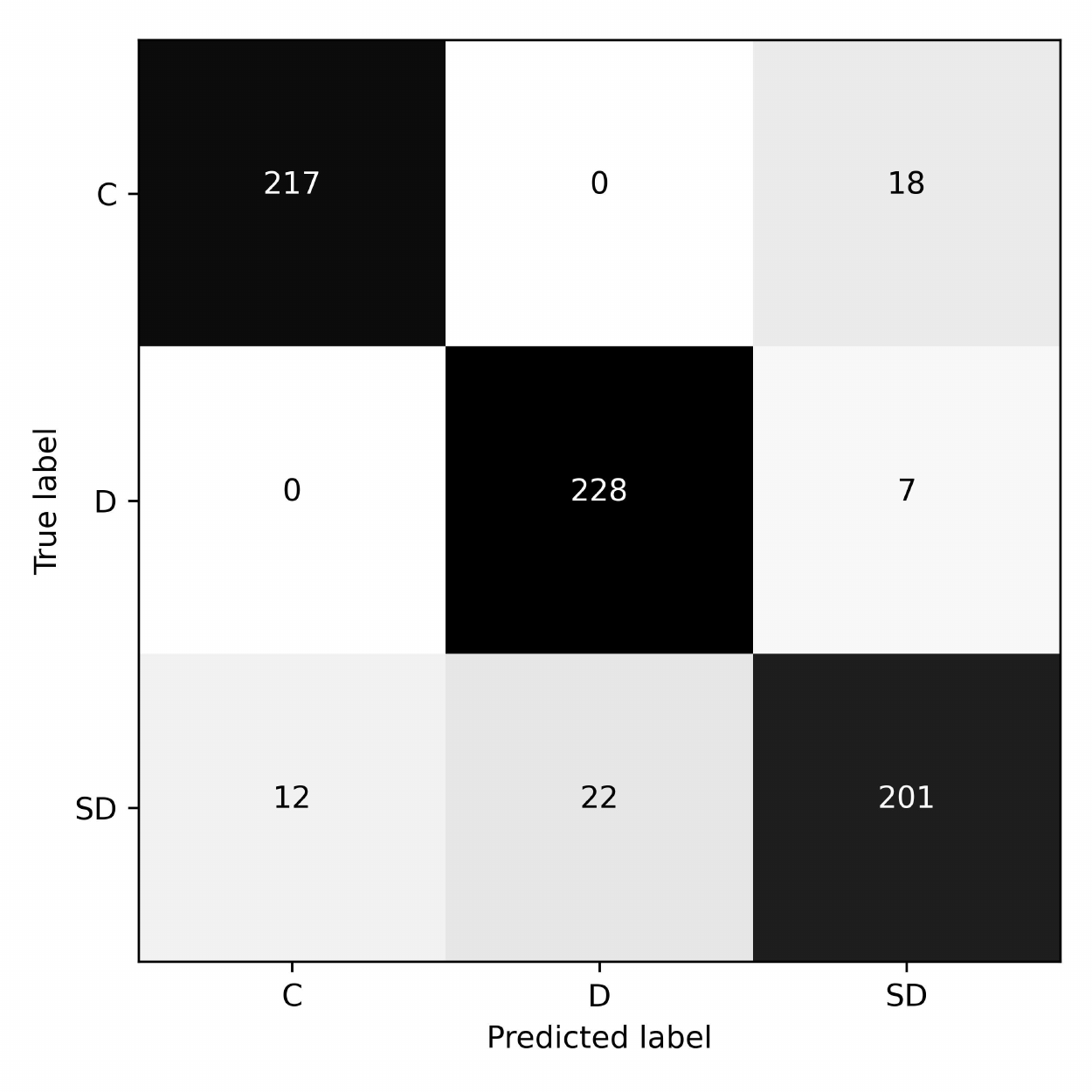}
\caption[]{Confusion matrix for validation dataset (705 images) as obtained by using {\it metrics} module of Scikit-learn \citep{ped11} based on Keras model file containing the final network architecture. C, D and SD refer to contact, detached and semi-detached morphologies, respectively.}
\label{conf_matrix}
\end{figure}

\begin{figure*}
\centering
\includegraphics[width=\columnwidth]{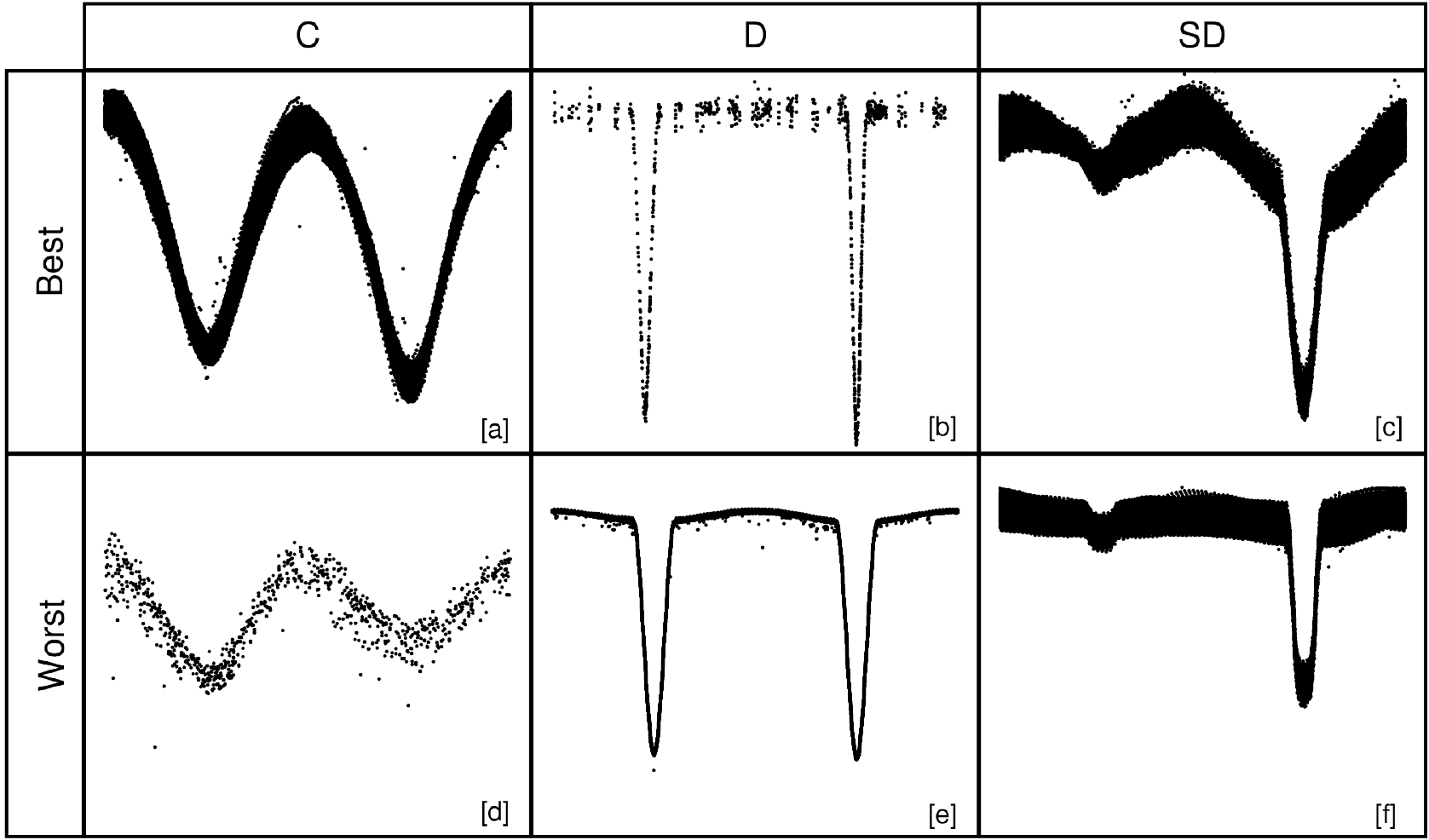}
\caption[]{Selection from the best and the worst classified light curves with their estimated morphological classes. The algorithm correctly classified KIC~10723143 (a), BW~Aqr (b) and KIC~06852488 (c) with the probability of 99.9\% as contact, detached and semi-detached, respectively. The semidetached binaries ASAS~125523-7322.2 (d) and KIC~10191056 (e) were estimated as contact and detached systems with 99.9\% probability. KIC~03530668 (e) was also misclassified as 93.1\% semidetached, while its actual class is detached. C, D and SD refer to contact, detached and semi-detached morphologies, respectively.}
\label{bestworst}
\end{figure*}

A plot of the confusion matrix (Fig.~\ref{conf_matrix}) for the validation dataset reveals the details of the classification result. 217 of contact, 228 of detached and 201 of semi-detached systems out of total 705 were correctly classified. The maximum misclassification was seen in semi-detached binaries; 22 of them are classified as detached systems. Selected light curves among the best and the worst classified data for each of the morphological classes are given in Fig.~\ref{bestworst}. True positive (TP), true negative (TN), false positive (FP) and false negative (FN) values calculated based on the confusion matrix are listed in Table~\ref{tpfp}.

\begin{table}
\centering
\normalsize
\begin{minipage}[]{100mm}
\caption[]{True positive ({\it TP}), true negative ({\it TN}), false positive ({\it FP}) and false negative ({\it FN}) values of classification for validation dataset covering 705 image data.} \label{tpfp}\end{minipage}
\setlength{\tabcolsep}{5pt}

\begin{tabular}{cccccc}
\hline
 & {\it TP} & {\it TN} & {\it FP} & {\it FN}  \\
 \hline
C & 217 & 458 & 12 & 18  \\
D & 228 & 448 & 22 & 7 \\
SD & 201 & 445 & 25 & 34 \\
\hline
\end{tabular}

\end{table}

The classification report, indicating metrics for the classification of the validation dataset, is shown in Table~\ref{class_rep}. Precision is the ratio of true positives to the total number of true and false positives ($ TP/ (TP+FP)$), a measure of how trustable is the model in predicting positive samples \citep{tin10}. Recall is defined as the ratio of the number of correctly classified positives to the total number of positives, $TP / (TP+FN)$, and it focuses on positive samples. F$_1$ score is the harmonic mean of precision and recall. In addition to these metrics, the subset accuracy of the classification, calculated using {\tt accuracy\_score} function of {{\tt scikitlearn}} library \citep{ped11}, is 92\%. This is simply the percentage of correctly classified samples \citep{tso07}:

\begin{equation}
     \frac{1} {|D|}\sum_{i=1}^{|D|} I(Z_i = Y_i)
\end{equation}
where $Y_i$ and $Z_i$ are actual and predicted labels, while $|D|$ is the number of multilabel examples and $I$ takes the value of 0 or 1 for false or true statements, respectively. 

\begin{table}
\centering
\normalsize
\begin{minipage}[]{100mm}
\caption[]{Classification report. C, D and SD refer to contact, detached and semi-detached systems, respectively.} \label{class_rep}\end{minipage}
\setlength{\tabcolsep}{4pt}

\begin{tabular}{ccccc}
\hline
 & Precision & Recall & F$_1$ score & number of data \\
 \hline
C & 0.948 & 0.923 & 0.935 & 235 \\
D & 0.912 & 0.970 & 0.940 & 235 \\
SD & 0.889 & 0.855 & 0.872 & 235 \\
\hline
Average & 0.916 & 0.916 & 0.916 & 705 \\
\hline
\end{tabular}

\end{table}

\begin{figure*}
\centering
\includegraphics[width=\columnwidth]{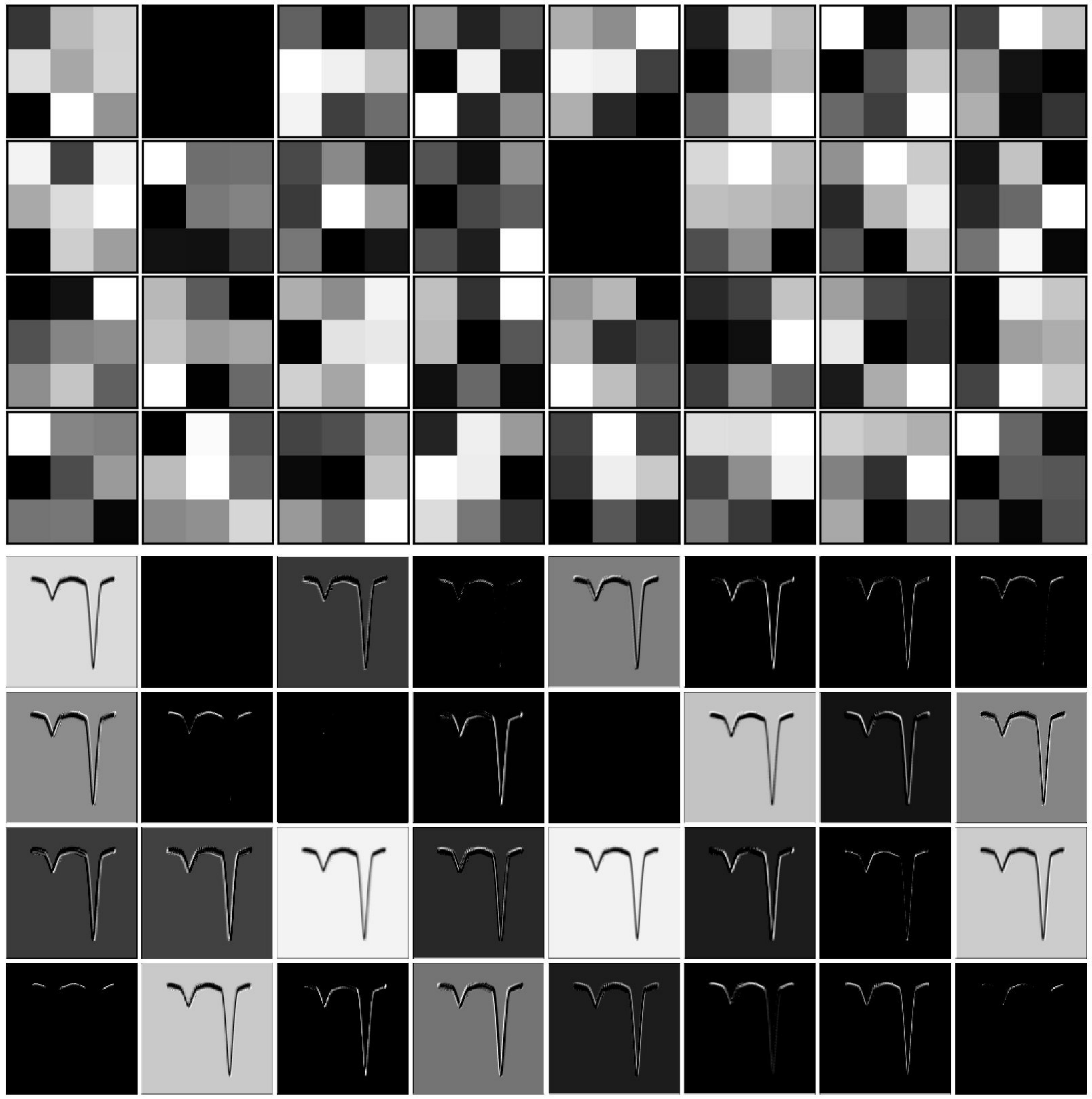}\\
\caption[]{32 $(3x3)$ filters of the first convolutional layer (upper panel) and 32 feature maps of the light curve image of KIC~03954798 as output from the first convolutional layer with corresponding filters (lower panel). Figure created using {\tt Matplotlib} \citep{hun07}.}
\label{conv_filt}
\end{figure*}
\begin{figure*}
\centering
\includegraphics[height=18cm, keepaspectratio,]{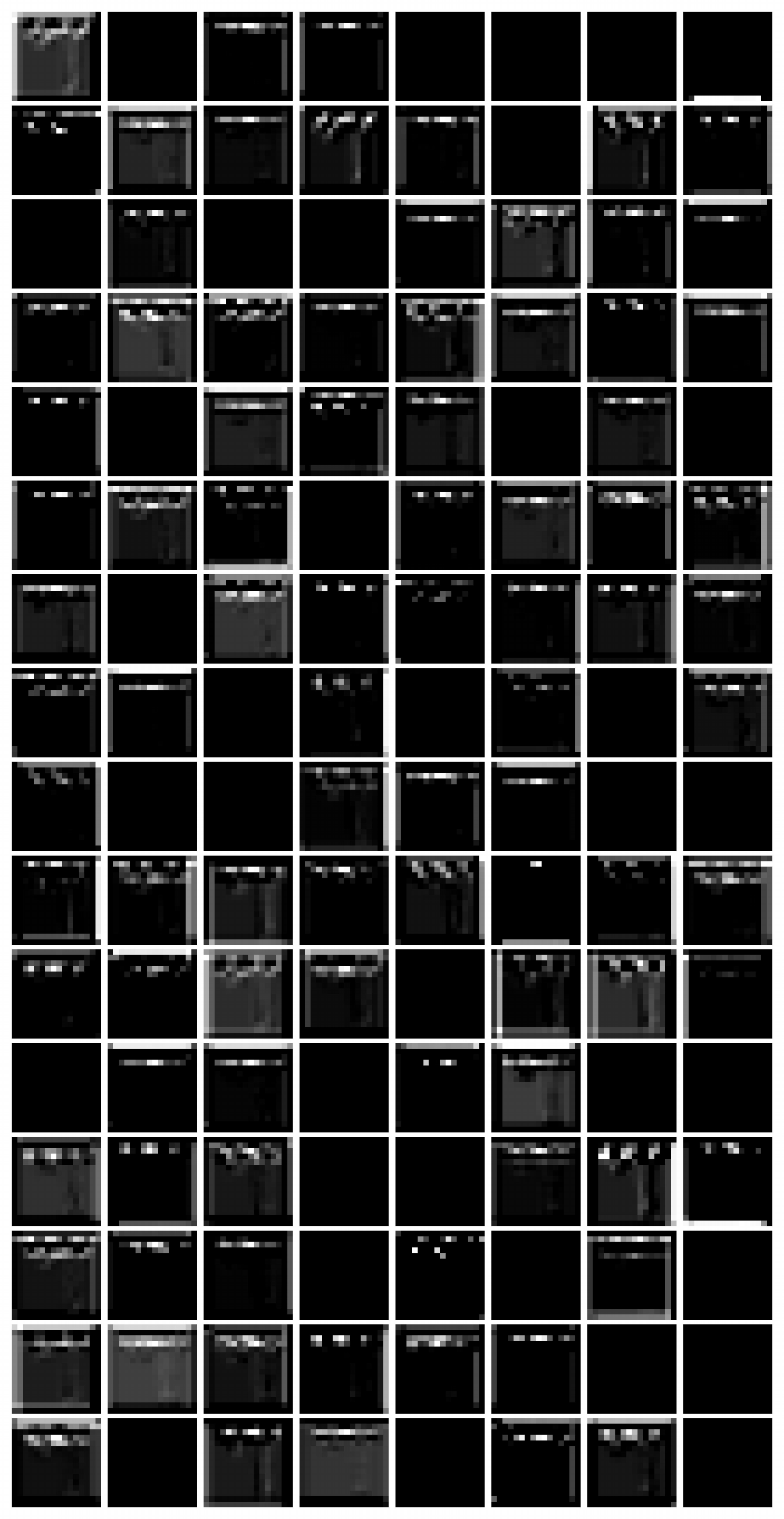}\\
\caption[]{Same as the lower panel of Fig.~\ref{conv_filt}, but for the fourth convolutional layer with 128 filters. Note that human perception for the light curve is almost lost.}
\label{conv_filt2}
\end{figure*}

Furthermore, it is worth looking up the filters and the output of convolutional layers (feature maps) of the final architecture in the way of how the machine sees and processes the light curves through the network. As an example, we demonstrate the feature maps for the light curve of KIC~03954798 as proceeded along with the filters of the first and the fourth convolutional layers in Fig.~\ref{conv_filt} and Fig.~\ref{conv_filt2}. For the human eye, the deeper the layer is, the harder the light curve perception is.

Beside fixing the seeds in our code for reproducibility, following the 2.0 version of the reproducibility checklist for machine learning given by \citet{pin20} we addressed the details of our model in the previous section. The algorithm was explained in detail with necessary mathematical descriptions. The application platform and infrastructure used were also denoted. The sample size of the data was given, the number of examples in the train and validation set was remarked, and the data preparation process was denoted (Sec.~\ref{lcdata}). The dataset and the code executing the classification are downloadable\footref{bu}. We defined and give the metrics of the classification and indicated the classification report, which refers to the quality of the classification.

The scientific importance of our neural network algorithm arises from its providing a way to distinguish the morphological types of eclipsing binary systems with high accuracy only using their light curve images. It is also a lot faster than other conventional methods conducting the same process, such as a workflow covering the trials of at least two morphologies in a widely known light curve analysis software and comparing the results to choose the best one. The determination of the morphological class is vital in the analysis of an eclipsing binary light curve in order to yield physically meaningful results, therefore, our algorithm can be applied to a light curve image before its analysis to establish a rapid and reliable morphological assumption for the light curve solution.

When it comes to comparing our results to other studies using machine learning algorithms related to the morphologies, our accuracy is found to be close to that of investigations in the literature. Although they were not to deal with the morphology alone, the accuracy in the three-layer artificial neural network by \citet{pri08}, which focused on detached morphological classification, was higher than 90\%. An image classification algorithm proposed by \citet{ula20} was also reached an accuracy value of 91\%. \citet{cok21a} achieved 98\% accuracy through their combined classifier, which was trained with synthetic light curve data constructed using ELISa software for detached and overcontact morphologies. We did not run our code using the images generated from the light curve of the above-mentioned studies, since a classification owes its resulting accuracy to properly collected training data as well as the architecture. A complete change in the training set most probably requires modification in the network architecture and hyperparameters to achieve the same accuracy, over 90\%.

The accurate information on the classes of training samples plays a vital role in the quality of the results. Therefore, as a future of the study, we plan to improve our algorithm by collecting light curve images with more accurate information on their types, namely the light curves of the systems having the morphological classes determined by analyses through human-controlled software, since hands-on modelling is the finest approach as \citet{koc20} concluded. This is projected to be done by a detailed survey of the literature for individual analyses of eclipsing binary light curves. Additionally, we publish a data collection platform\footref{bu} to where the researchers from the community can upload morphological information and light curve data of human-confirmed eclipsing binary stars. Thuswise, the number of training and validation samples, another crucial parameter, is also aimed to be increased. The increasing number of space telescope data of binary stars will boost the number of samples without a doubt, as long as morphological classes are precisely determined. Finally, our code and collected data are public, therefore, it is open to be improved by tuning the hyperparameters or altering the architecture by the researchers from the area.

\section*{Acknowledgements}

This paper includes data collected by the Kepler mission and obtained from the MAST data archive at the Space Telescope Science Institute (STScI). Funding for the Kepler mission is provided by the NASA Science Mission Directorate. STScI is operated by the Association of Universities for Research in Astronomy, Inc., under NASA contract NAS 5–26555. 

\section*{Data Availability}

The datasets generated and analyzed during the current study are available in the GitHub repository, \url{https://github.com/burakulas/ebclass}.

\bibliographystyle{plainnat2}
\bibliography{Ulas_bib}

\end{document}